\documentclass[aip,apl,reprint]{revtex4-1}

\usepackage{graphicx}
\usepackage{dcolumn}
\usepackage{bm}
\usepackage{siunitx}

\begin{document}

\title{Effect of oxygen plasma on nanomechanical silicon nitride  resonators}

\author{Niklas Luhmann}
\affiliation{%
Institute of Sensor and Actuator Systems, TU Wien, Vienna, Austria
}
\affiliation{%
Department of Physics, University of Konstanz, 78457 Konstanz, Germany
}
\author{Artur Jachimowicz}
\author{Johannes Schalko}
\author{Pedram Sadeghi}
\affiliation{%
Institute of Sensor and Actuator Systems, TU Wien, Vienna, Austria
}
\author{Markus Sauer}
\author{Annette Foelske-Schmitz}
\affiliation{%
Analytical Instrumentation Center, TU Wien, Vienna, Austria
}
\author{Silvan Schmid}%
 \email{silvan.schmid@tuwien.ac.at}
\affiliation{%
Institute of Sensor and Actuator Systems, TU Wien, Vienna, Austria
}%

\date{\today}

\begin{abstract}
Precise control of tensile stress and intrinsic damping is crucial for the optimal design of nanomechanical systems for sensor applications and quantum optomechanics in particular.
In this letter we study the influence of oxygen plasma on the tensile stress and intrinsic damping of nanomechanical silicon nitride resonators. Oxygen plasma treatments are common steps in micro
and nanofabrication. We show that oxygen plasma of only a few minutes oxidizes the silicon nitride surface, creating several nanometer thick silicon dioxide layers with a compressive stress of \SI{1.30+-0.16}{\giga\pascal}. Such oxide layers can cause a reduction of the effective tensile stress of a \SI{50}{\nano\meter} thick stoichiometric silicon nitride membrane by almost 50\%. Additionally, intrinsic damping linearly increases with the silicon dioxide film thickness. An  oxide layer of \SI{1.5}{\nano\meter}  grown in just \SI{10}{\second} in a  \SI{50}{\watt}  oxygen plasma almost doubled the intrinsic damping. The oxide surface layer can be efficiently removed in buffered HF.    
\end{abstract}

\maketitle

\section{\label{sec:introduction}Introduction}
Silicon nitride  has become a much valued material for the fabrication of nanomechanical resonators due to its excellent mechanical and optical properties. A particularly interesting feature of silicon nitride thin films  is the large intrinsic tensile stress. This stress   not only defines the resonance frequency ($f$) of  resonators such as strings or membranes but  further  dilutes  intrinsic damping mechanisms, \cite{Gonzfilez1994,schmid2008damping,Schmid2011,Yu2012,Schmid2016} which results in exceptionally high quality factors ($Q$). \cite{verbridge2006high,zwickl2008high,Verbridge2008,Schmid2011,Chakram2014}
This has made nanomechanical silicon nitride  resonators 
a favorite choice e.g. for   cavity optomechanics experiments. \cite{Purdy2013,Wilson2012,Gavartin2012,Bagci2014,thompson2008strong,Brawley2014a}

In particular for applications in quantum  optomechanics there are strong efforts underway in order to overcome the theoretically required limit of  $Q\times f >$ \SI{1e13}{\hertz,} which would enable  quantum experiments at room temperature.
There are basically two approaches on how to improve $Q$, which are  by i) optimizing  damping dilution either by "soft-clamping" of silicon nitride resonators inside phononic crystal structures,
\cite{Tsaturyan2016,Ghadimi2016} or by increasing the tensile stress,\cite{Norte2015} and ii)
by reducing intrinsic losses.

More generally, the precise control of mechanical parameters, such as tensile stress and intrinsic damping, are  of fundamental significance for the optimal design of nanomechanical sensors.
In particular, the responsivity of spectrochemical sensors based on the photothermal heating of a silicon nitride resonator directly depends on the magnitude of tensile stress.\cite{Kurek2017,Andersen2016,Yamada2013b,Biswas2014,Larsen2013}

In this letter we study the effect of oxygen plasma on both  effective tensile stress and  intrinsic loss of nanomechanical silicon nitride  membrane resonators.  The incineration of polymeric photoresist residues with an oxygen plasma
is common
practice in nano and microfabrication.
Although oxygen plasma has long been known to not only effectively oxidize
silicon \cite{Ligenza1965,Kraitchman1967,Pulfrey1974} but also  silicon nitride,
\cite{Taylor1988,Kennedy1995,Kennedy1999} its effect on  nanomechanical resonators
 has so far not been recognized.

\section{\label{sec:methods}Methods}
The experiments were done with rectangular silicon-rich (low-stress) and stoichiometric (high-stress) silicon nitride membranes. The membranes were fabricated from Si wafers coated with \SI{50}{\nano\meter}  silicon nitride by low-pressure chemical vapor deposition (LPCVD), purchased from Hahn-Schickard-Gesellschaft f\"ur angewandte Forschung e.V. with a nominal tensile stress of $\sigma_0 \approx$ \SI{50}{\mega\Pa} and $\sigma_0 \approx$ \SI{1}{GPa}, respectively.
The membranes were patterned by photolithography and dry etching of the backside silicon nitride layer and subsequently released by anisotropic KOH (\SI{40}{wt\%})\ wet etching all through the silicon wafer. 

The oxygen plasma exposure was performed with a parallel plate STS320PC RIE plasma  system from STS Systems with \SI{49,5}{sccm} O$_2$ flow and a chamber pressure of \SI{20}{\pascal}.

The vibrational analysis of the membranes was conducted under high vacuum  with a laser-Doppler vibrometer (MSA-500 from Polytec GmbH). The membranes were actuated thermoelastically by focusing an amplitude modulated diode laser ($\lambda=$\SI{635}{\nano\meter}, with a maximal power on the sample of \SI{70}{\micro\watt}) onto the membrane rim.

The quality factors $Q$ were extracted from ring-down measurements performed with a lock-in amplifier (HF2LI from Zurich Instrument). The ring-down was prepared by first driving the specific  resonance mode with a phase-locked loop before stopping the actuation.

The EDX analysis was performed with a
\SI{20}{\micro\meter} area scan directly on the membrane using
an X-Max$^\text{N}$ detector provided by Oxford Instruments attached to
a Hitachi SU8030 scanning electron microscope.

The XPS measurements were performed with an SPECS XPS-spectrometer, equipped with  a monochromatic  aluminium K-alpha X-ray source ($\mu$Focus 350) and a hemispheric WAL-150 analyser. Additional sample preparation was carried out using 3kV Ar+-ions from a SPECS IQ 12/38 ion sputter gun.
The surface composition analysis was supported by simulation of electron spectra for surface analysis (SESSA) software.\cite{Smekal2005} 

The thin film thickness was measured with a Filmetrics F20-UVX thin film analyzer.

\begin{figure}
\begin{center}
\includegraphics[width=0.4\textwidth]{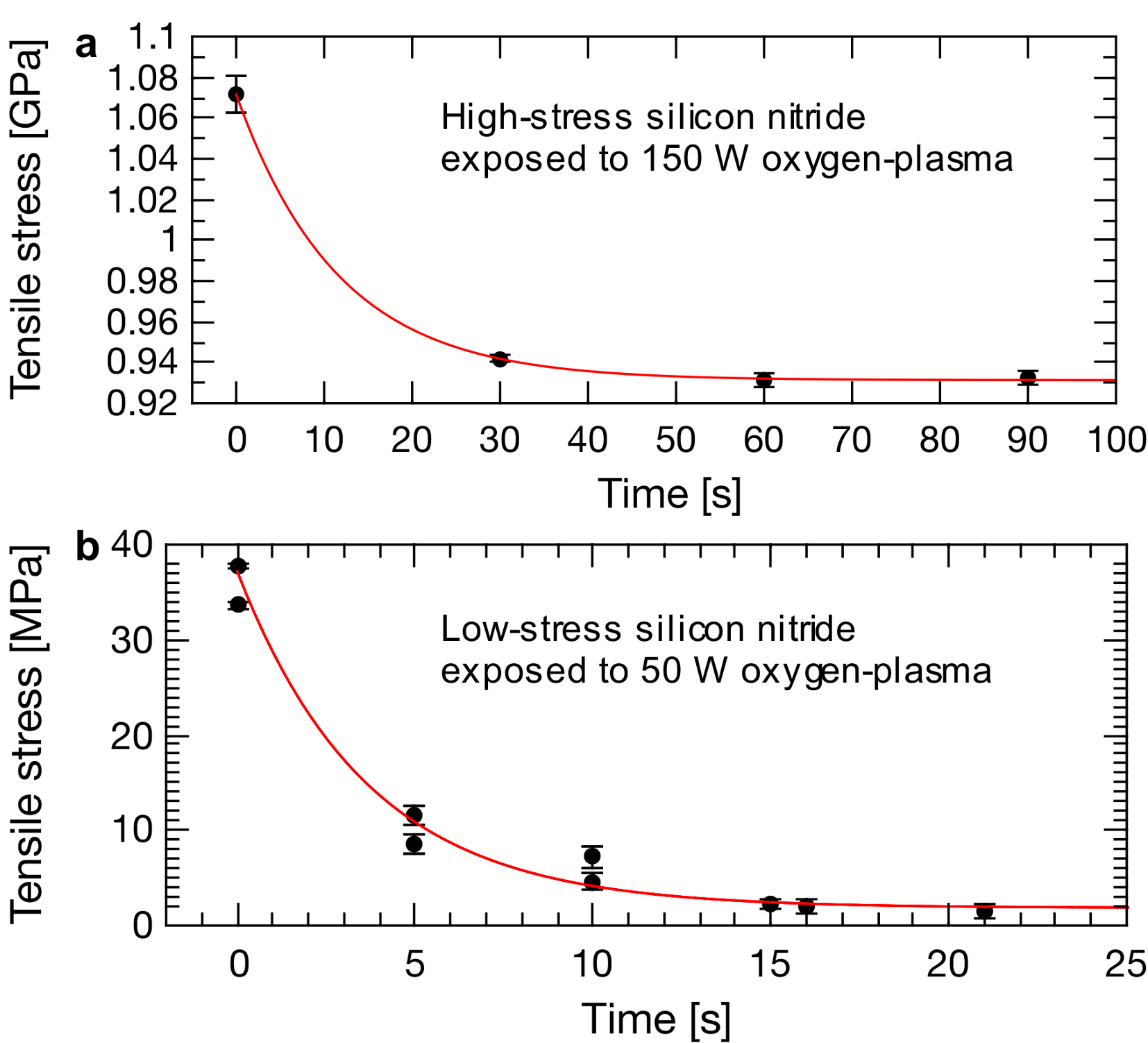}
\caption{\label{fig:1}Tensile stress of a) high-stress   ($L=$\SI{500}{\micro\meter})
and b) low-stress silicon nitride membranes ($L=$\SI{500}{\micro\meter}) with respect to the 
 oxygen plasma exposure time. Each stress value  is the average
of 5 membranes extracted from the fundamental mode (1,1). The red lines are exponential fits. }
\end{center}
\end{figure}
\section{\label{sec:RandD}Results \& Discussion}
Figure~\ref{fig:1} shows the tensile stress of high-stress and low-stress silicon nitride\ membranes for an increasing  time in oxygen plasma. The tensile stress $\sigma$ was extracted from the eigenfrequency $f_{n,m}$
model for membranes \cite{Schmid2016}
\begin{equation}\label{eq:membrane}
f_{n,m}=\frac{\sqrt{n^2+m^2}}{2L}\sqrt{\frac{\sigma}{\rho}},
\end{equation}
with length $L$, mode numbers $n$ and $m$, and assuming a mass density of
$\rho=$\SI{3000}{kg/m^3}.
Comparisons with an extended plate model have shown that the ideal membrane model (\ref{eq:membrane}) holds true for all analysed membranes under tensile stress. For both types of membranes, tensile stress drops exponentially with plasma
exposure time. The intrinsic tensile stress in the high-stress silicon nitride membranes dropped by almost \SI{150}{\mega\pascal} when shortly exposed to an  oxygen plasma of \SI{150}{\watt} (see Figure~1a). For low-stress silicon nitride membranes the initial \SI{40}{\mega\pascal} tensile stress reduced to almost zero for an exposure to only \SI{50}{\watt} (see Figure~1b).  When treating the low-stress silicon nitride with \SI{150}{\watt} oxygen plasma  the stress even reversed from tensile to compressive, as can be seen from microscope images shown  in  Figure~2a.\begin{figure}
\begin{center}
\includegraphics[width=0.44\textwidth]{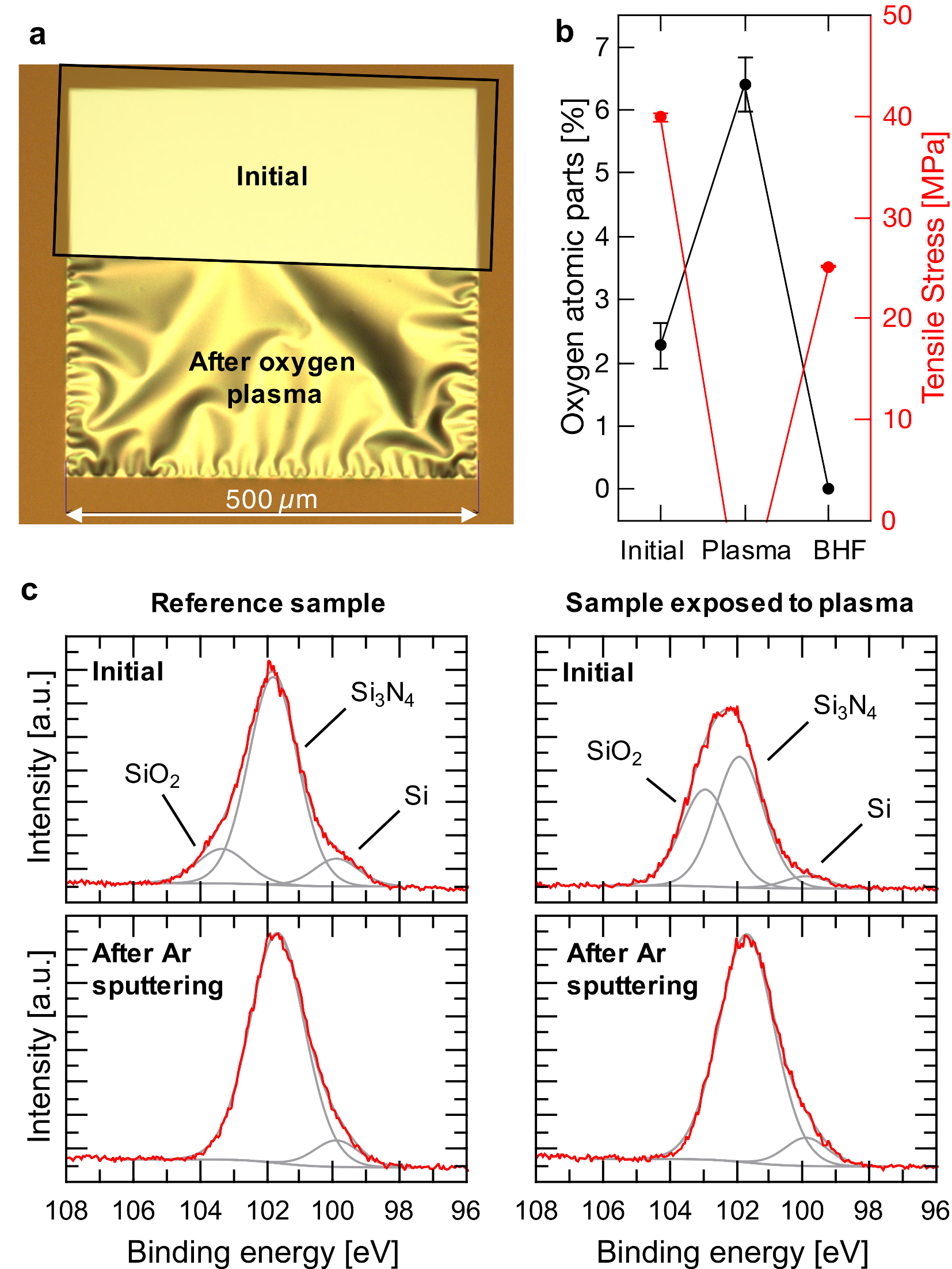}
\caption{\label{fig:2}a) Microscope images of a low-stress silicon nitride membrane before and after \SI{30}{\second} exposure to \SI{150}{\watt} oxygen plasma.
b) EDX analysis of the atomic composition and tensile
stress of a low-stress silicon nitride\ membrane ($L=500$~$\mu$m) measured initially,
after \SI{30}{\second}  of \SI{150}{\watt} oxygen plasma, and after \SI{2}{\minute}  in BHF.  Each data point represents the average
of 5 membranes. c) Normalized XPS detail spectra of low-stress silicon nitride samples, measured initially
and after  \SI{30}{\s} of Ar-ion sputtering. The sample treated by oxygen
plasma was exposed for \SI{21}{\s} at \SI{50}{\watt}.   }
\end{center}
\end{figure}

Figure~2b presents the atomic composition and tensile stress of low-stress silicon nitride membranes measured i) initially, ii) after a \SI{30}{\second} exposure to \SI{150}{\watt} oxygen plasma, and iii) after a \SI{2}{\min} bath in buffered hydrofluoric acid (BHF). As mentioned before, \SI{150}{\watt} oxygen plasma resulted in compressive stress, clearly visible by the ripples in the membrane in Figure~\ref{fig:2}a, and an increased oxygen content. A subsequent dip in buffered HF (BHF)  recovered the tensile stress. The same recovery was also found for high-stress silicon nitride membranes, whose stress reached 96\% of the initial value after a \SI{1}{\min} dip in BHF (data not shown).  This is clear evidence that the stress reduction is caused by a  surface layer with compressive stress and that  is removable in BHF. It is known that oxygen plasma not only  creates a compressive silicon dioxide layer in silicon substrate,\cite{Pulfrey1974} but it also efficiently oxidises LPCVD silicon nitride thin films.\cite{Taylor1988} The rise in atomic oxygen content after plasma seen in Figure~\ref{fig:2}b  can be attributed to the creation of such an oxidation layer.

From previous work it appears that the oxygen from the plasma substitutes the nitrogen in the silicon nitride film. In detail, these oxygen plasma grown silicon dioxide films  feature a vertical  gradient in composition ranging from  i) a  silicon dioxide layer at the surface, due to a loss of all nitrogen atoms into the plasma, to ii) pure silicon nitride at the interface.  \cite{Buiu1998,Kennedy1999,Kennedy1995} Evidence of the growth of such a silicon dioxide layer was obtained by  XPS  analysis, which revealed that the surface of the plasma induced oxidation
layer indeed is fully depleted of nitrogen. Figure~\ref{fig:2}c presents the chemical
state analysis of the Si 2p region a silicon nitride sample before (reference sample) and after oxygen
plasma exposure. Deconvolution of these Si 2p peaks shows that both samples exhibit an SiO$_2$ (103.5 eV) component in addition to Si$_3$N$_4$ (101.7 eV). However, the plasma treated silicon nitride shows a significant  SiO$_2$-related component, representing a substantial SiO$_2$ film,  with a thickness of several nm. In contrast  to the faint SiO$_2$ signature of the reference sample, which can be assigned to a few monolayers of native SiO$_2$ at the silicon nitride surface.   After sputtering of the sample surface with Ar ions, the SiO$_2$ related peaks disappeared completely for both samples.

In the plasma oxidation process, the volume of  Si$_3$N$_4$ molecules of 77.5~\AA$^3$ (calculated assuming a  mass density of 3000~kg/m$^3$) grows to the volume of three SiO$_2$ molecules of 45.5~\AA$^3$ each (assuming a mass density of 2200~kg/m$^3$). In the case of planar growth the consumed Si$_3$N$_4$ is replaced by a SiO$_2$ film that is theoretically 1.77$\times$  thicker, which has been verified with electron microscope images of oxidised silicon nitride surfaces.\cite{Taylor1988} The growth of  SiO$_2$  leads to a one-dimensional unrelaxed strain of $\varepsilon=0.21$, which would result in an enormous compressive stress of $\sigma_{SiO}=\varepsilon E/(1-\nu) \approx$ \SI{18}{\giga\Pa}. This is of the same order of magnitude of the initial stress building up  during silicon  oxidation.\cite{Sutardja1989}
However, this enormous compressive stress is expected to relax to a magnitude close to the compressive strength of amorphous silicon dioxide of  approximately \SI{1.1}{\giga\Pa}.

\begin{figure}
\begin{center}
\includegraphics[width=0.4\textwidth]{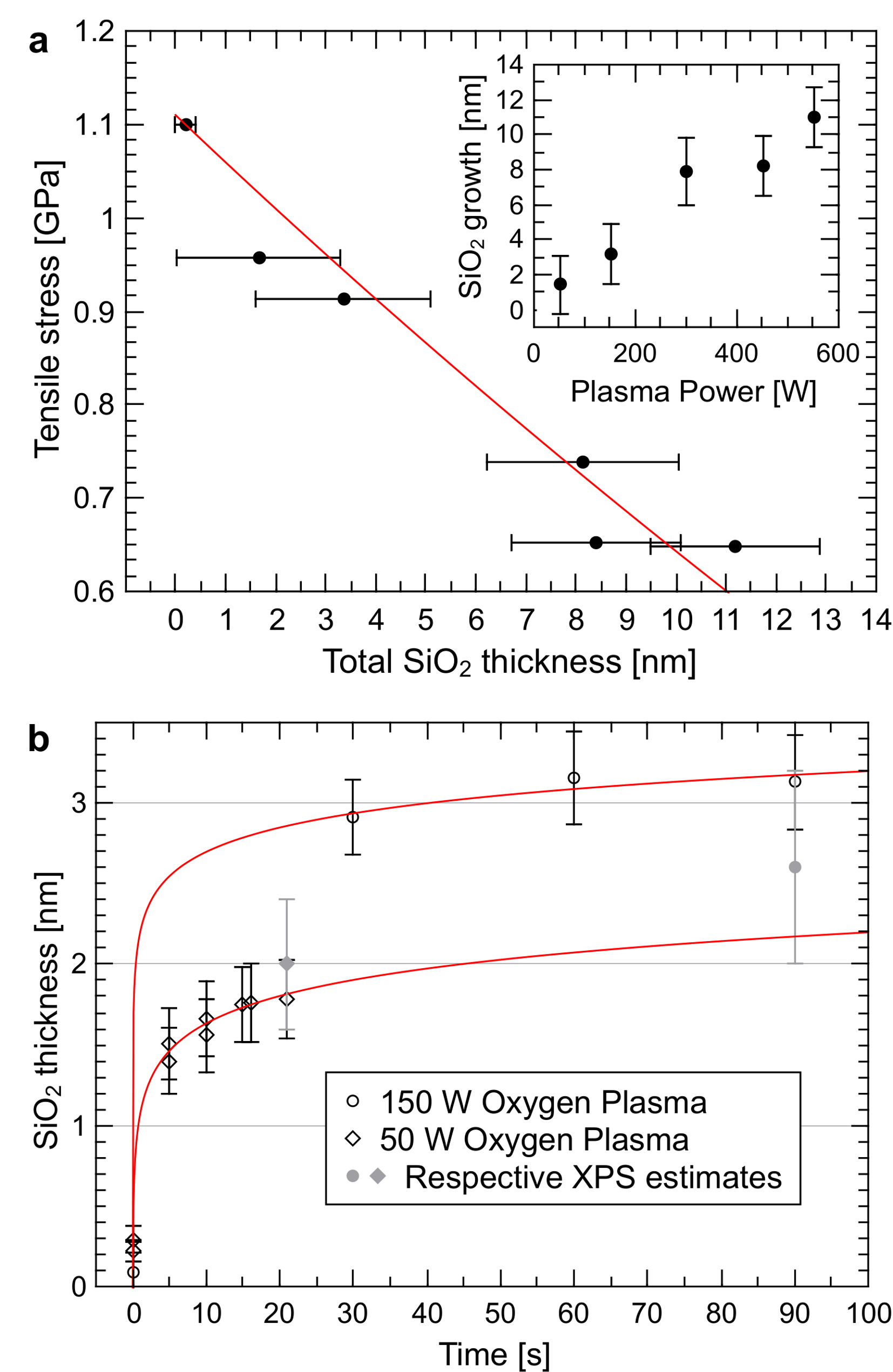}
\caption{\label{fig:3}a) Stress reduction of high-stress silicon nitride
membranes ($L=$\SI{500}{\micro\meter}) versus plasma power. The samples were
kept in the oxygen plasma for \SI{5}{\min}. The silicon dioxide film thickness
was estimated from (\ref{eq:meanstress}). The inset shows the measured silicon dioxide film thickness as a function of oxygen plasma power for an exposure time of 5 min each.  b) Effective SiO$_2$ film thickness
versus plasma exposure
time, calculated from (\ref{eq:meanstress}) based on the measured tensile
stress $\sigma$ of high-stress and low-stress silicon nitride, as presented
in Figure~\ref{fig:1}. }
\end{center}
\end{figure}

In order to estimate the effective compressive stress of the silicon dioxide layer, high-stress silicon nitride membranes were oxidised with various plasma powers. Afterward the grown SiO$_2$  was removed in BHF. The  SiO$_2$ layer thickness was then calculated  from the reduction of the silicon nitride film thickness (see inset of Figure~\ref{fig:3}a), taking into account the volume increase during oxidation and the BHF etch-rate of the silicon nitride. Figure~\ref{fig:3}a shows the tensile stress $\sigma$, obtained from measured membrane resonance frequencies, versus the measured oxide film thickness. The effective stress $\sigma$ in the silicon nitride membrane can be modelled as the arithmetic mean of the tensile stress $\sigma_{SiN}$ of the silicon nitride (of thickness   $h_{SiN}$) and the compressive stress $\sigma_{SiO}$ of the silicon dioxide layer (of thickness $h_{SiO}$)\begin{equation}\label{eq:meanstress}
\sigma =   \frac{\sigma_{SiO}h_{SiO} + \sigma_{SiN}h_{SiN}}{h_{SiO}+h_{SiN}}.
\end{equation}
Fitting  the model (\ref{eq:meanstress}) to the data in Figure~\ref{fig:2}a allows the extraction of a compressive stress of the SiO$_2$ film of \SI{1.30+-0.16}{\giga\pascal}. The measured stress is of
the expected magnitude of the compressive strength of amorphous silicon dioxide.

With the gained value for the compressive stress  $\sigma_{SiO}$, it is now possible to estimate the oxide film thickness based on the measured effective stress $\sigma$.   Figure~\ref{fig:3}b plots the estimated SiO$_2$  thickness of the silicon dioxide layer as a function of oxygen plasma time for the two samples  presented in Figure~\ref{fig:1}. The same samples have been analysed with XPS  and quantitative results have subsequently been compared with simulations varying the SiO$_2$ layer thickness, and the obtained thickness estimates match well with the estimated values  from the effective tensile stress by means of  (\ref{eq:meanstress}). It has been shown that oxygen plasma induced  oxide growth at room temperature shows logarithmic behavior.\cite{Kim1996} The same behavior seems to hold true for the oxygen plasma grown from silicon dioxide thin films, as can be seen by the logarithmic fits shown as red lines.

\begin{figure}
\begin{center}
\includegraphics[width=0.4\textwidth]{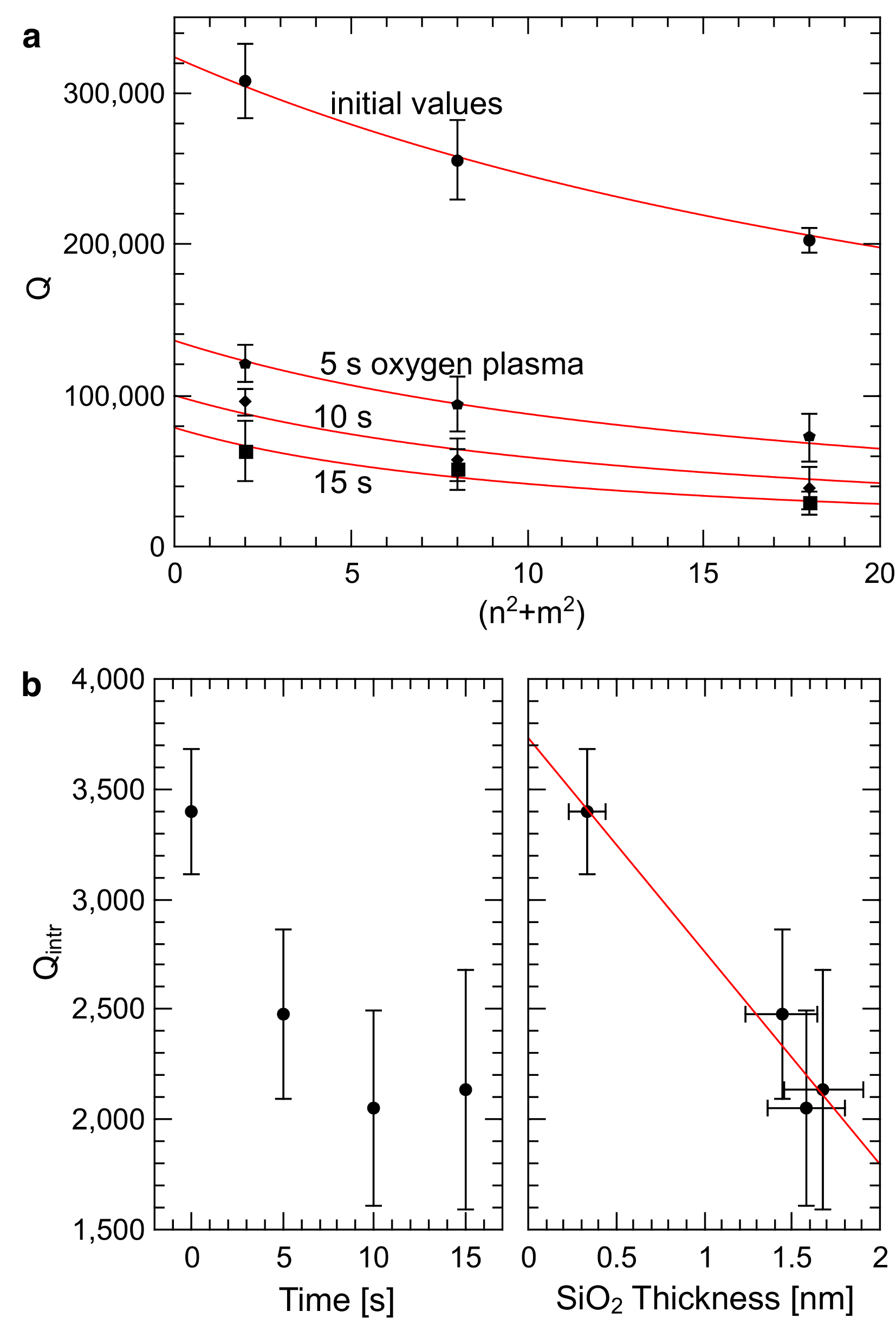}
\caption{\label{fig:4} Energy loss study of 5 low-stress silicon nitride membranes ($L=$\SI{200}{\micro\meter}) for the (1,1), (2,2), and (3,3) mode after different exposure times to \SI{50}{\W} oxygen plasma. The red lines represent fits with the damping  dilution model (\ref{eq:dd}) with $Q_{intr}$ as fitting parameter. b) Extracted average $Q_{intr}$ values for the different plasma exposure times. The intrinsic quality factor values are further plotted versus the silicon dioxide thickness estimated from the measured effective stress. The red line is a linear fit.}
\end{center}
\end{figure}

In order to study the effect of oxygen plasma on intrinsic losses, the quality factor of low-stress silicon nitride membranes was measured, as presented in Figure~\ref{fig:4}a. The low tensile stress in these membranes produce a sufficient decoupling from the chip frame thereby minimizing radiation losses.\cite{Villanueva2014b} Hence,   the measured    quality factors are exclusively limited by intrinsic damping, which is further confirmed by the fact that the measured values for each treatment step follow the prediction (red curves) from the damping dilution model (\ref{eq:dd}) for intrinsic loss (1/$Q_{intr}$) in membranes\cite{Yu2012, Schmid2016}
\begin{equation}\label{eq:dd}
Q=\left[\frac{h}{L}\sqrt{\frac{E}{3\sigma}} + \frac{\pi^2(n^2+m^2)}{12}\frac{
E}{\sigma} \frac{h^2}{L^2}  \right]^{-1} Q_{intr}.
\end{equation}
The intrinsic loss (1/$Q_{intr}$) was then extracted  by correcting the
measured quality factor values $Q$ for the stress induced damping dilution
effect  (\ref{eq:dd}) 
assuming a Young's modulus of $E=$ \SI{200}{\GPa} for silicon-rich silicon
nitride.\cite{Guo2003}
The respective intrinsic quality factors $Q_{intr}$ for the different plasma exposure times are plotted in Figure~\ref{fig:4}b, together with the correlation to the estimated silicon dioxide film thickness. Apparently, intrinsic losses increase with oxygen plasma exposure.  The comparison of $Q_{intr}$ with the estimated  oxide thickness shows a  linear correlation. This  suggests that the increased loss can be attributed to the growth of the silicon dioxide surface layer, which has larger intrinsic damping than silicon nitride.\cite{southworth2009stress}  The magnitude of the initial  $Q_{intr}$ value matches the general value found for surface loss in silicon nitride.\cite{Villanueva2014b}

\section{\label{sec:conclusion}Conclusion}
Silicon nitride structures  with a  thickness of only \SI{50}{\nano\meter}
 are highly sensitive to oxygen plasma. Even short exposures cause a significant
 decreases in   tensile stress and  an increase in intrinsic loss. Both effects can be attributed to the  plasma-induced oxidation of the silicon nitride
surface. We found that the created silicon dioxide  film on the silicon nitride surface has a compressive stress of \SI{1.30+-0.16}{\giga\pascal}, which is probably limited by the layers' own compressive strength.

This relatively large stress counteracts the intrinsic tensile stress in LPCVD silicon nitride, leading to a significant drop of the intrinsic stress and hence in the resonance frequency. Oxygen plasma exposure of nanomechanical silicon nitride resonators can reduce the tensile stress by several hundreds of MPa, which for low-stress silicon nitride structures can even cause a total reversal of the effective stress from tensile to compressive. Hence it is an interesting tool which allows a precise post-fabrication control of tensile stress.

Additionally, the plasma grown silicon dioxide layer significantly increases energy loss in nanomechanical silicon nitride resonators. Hence, for applications where a maximum tensile stress as well as minimum intrinsic damping is desired, the silicon dioxide layer and its deteriorating effects can fully be removed by a quick BHF dip.

Since an oxygen plasma also oxidizes silicon surfaces, similar effects as observed for silicon nitride resonators will likely occur in nanomechanical structures made from silicon.

\begin{acknowledgments}
We thank Sophia Ewert and Patrick Meyer for the   cleanroom support.
This project has received funding from the European Research Council (ERC) under the European Union's Horizon 2020 research and innovation program (grant agreement - 716087 - PLASMECS).
\end{acknowledgments}


\begin{thebibliography}{36}%
\makeatletter
\providecommand \@ifxundefined [1]{%
 \@ifx{#1\undefined}
}%
\providecommand \@ifnum [1]{%
 \ifnum #1\expandafter \@firstoftwo
 \else \expandafter \@secondoftwo
 \fi
}%
\providecommand \@ifx [1]{%
 \ifx #1\expandafter \@firstoftwo
 \else \expandafter \@secondoftwo
 \fi
}%
\providecommand \natexlab [1]{#1}%
\providecommand \enquote  [1]{``#1''}%
\providecommand \bibnamefont  [1]{#1}%
\providecommand \bibfnamefont [1]{#1}%
\providecommand \citenamefont [1]{#1}%
\providecommand \href@noop [0]{\@secondoftwo}%
\providecommand \href [0]{\begingroup \@sanitize@url \@href}%
\providecommand \@href[1]{\@@startlink{#1}\@@href}%
\providecommand \@@href[1]{\endgroup#1\@@endlink}%
\providecommand \@sanitize@url [0]{\catcode `\\12\catcode `\$12\catcode
  `\&12\catcode `\#12\catcode `\^12\catcode `\_12\catcode `\%12\relax}%
\providecommand \@@startlink[1]{}%
\providecommand \@@endlink[0]{}%
\providecommand \url  [0]{\begingroup\@sanitize@url \@url }%
\providecommand \@url [1]{\endgroup\@href {#1}{\urlprefix }}%
\providecommand \urlprefix  [0]{URL }%
\providecommand \Eprint [0]{\href }%
\providecommand \doibase [0]{http://dx.doi.org/}%
\providecommand \selectlanguage [0]{\@gobble}%
\providecommand \bibinfo  [0]{\@secondoftwo}%
\providecommand \bibfield  [0]{\@secondoftwo}%
\providecommand \translation [1]{[#1]}%
\providecommand \BibitemOpen [0]{}%
\providecommand \bibitemStop [0]{}%
\providecommand \bibitemNoStop [0]{.\EOS\space}%
\providecommand \EOS [0]{\spacefactor3000\relax}%
\providecommand \BibitemShut  [1]{\csname bibitem#1\endcsname}%
\let\auto@bib@innerbib\@empty
\bibitem [{\citenamefont {Gonzfilez}\ and\ \citenamefont
  {Saulson}(1994)}]{Gonzfilez1994}%
  \BibitemOpen
  \bibfield  {author} {\bibinfo {author} {\bibfnamefont {G.~I.}\ \bibnamefont
  {Gonzfilez}}\ and\ \bibinfo {author} {\bibfnamefont {P.~R.}\ \bibnamefont
  {Saulson}},\ }\href@noop {} {\bibfield  {journal} {\bibinfo  {journal}
  {Journal of the Acoustical Society of America}\ }\textbf {\bibinfo {volume}
  {96}},\ \bibinfo {pages} {207} (\bibinfo {year} {1994})}\BibitemShut
  {NoStop}%
\bibitem [{\citenamefont {Schmid}\ and\ \citenamefont
  {Hierold}(2008)}]{schmid2008damping}%
  \BibitemOpen
  \bibfield  {author} {\bibinfo {author} {\bibfnamefont {S.}~\bibnamefont
  {Schmid}}\ and\ \bibinfo {author} {\bibfnamefont {C.}~\bibnamefont
  {Hierold}},\ }\href {\doibase 10.1063/1.3008032} {\bibfield  {journal}
  {\bibinfo  {journal} {Journal of Applied Physics}\ }\textbf {\bibinfo
  {volume} {104}},\ \bibinfo {pages} {093516} (\bibinfo {year}
  {2008})}\BibitemShut {NoStop}%
\bibitem [{\citenamefont {Schmid}\ \emph {et~al.}(2011)\citenamefont {Schmid},
  \citenamefont {Jensen}, \citenamefont {Nielsen},\ and\ \citenamefont
  {Boisen}}]{Schmid2011}%
  \BibitemOpen
  \bibfield  {author} {\bibinfo {author} {\bibfnamefont {S.}~\bibnamefont
  {Schmid}}, \bibinfo {author} {\bibfnamefont {K.~D.}\ \bibnamefont {Jensen}},
  \bibinfo {author} {\bibfnamefont {K.~H.}\ \bibnamefont {Nielsen}}, \ and\
  \bibinfo {author} {\bibfnamefont {A.}~\bibnamefont {Boisen}},\ }\href
  {\doibase 10.1103/PhysRevB.84.165307} {\bibfield  {journal} {\bibinfo
  {journal} {Physical Review B}\ }\textbf {\bibinfo {volume} {84}},\ \bibinfo
  {pages} {165307} (\bibinfo {year} {2011})}\BibitemShut {NoStop}%
\bibitem [{\citenamefont {Yu}, \citenamefont {Purdy},\ and\ \citenamefont
  {Regal}(2012)}]{Yu2012}%
  \BibitemOpen
  \bibfield  {author} {\bibinfo {author} {\bibfnamefont {P.-L.}\ \bibnamefont
  {Yu}}, \bibinfo {author} {\bibfnamefont {T.}~\bibnamefont {Purdy}}, \ and\
  \bibinfo {author} {\bibfnamefont {C.~A.}\ \bibnamefont {Regal}},\ }\href
  {\doibase 10.1103/PhysRevLett.108.083603} {\bibfield  {journal} {\bibinfo
  {journal} {Physical Review Letters}\ }\textbf {\bibinfo {volume} {108}},\
  \bibinfo {pages} {083603} (\bibinfo {year} {2012})}\BibitemShut {NoStop}%
\bibitem [{\citenamefont {Schmid}, \citenamefont {Villanueva},\ and\
  \citenamefont {Roukes}(2016)}]{Schmid2016}%
  \BibitemOpen
  \bibfield  {author} {\bibinfo {author} {\bibfnamefont {S.}~\bibnamefont
  {Schmid}}, \bibinfo {author} {\bibfnamefont {L.~G.}\ \bibnamefont
  {Villanueva}}, \ and\ \bibinfo {author} {\bibfnamefont {M.~L.}\ \bibnamefont
  {Roukes}},\ }\href {\doibase 10.1007/978-3-319-28691-4} {\emph {\bibinfo
  {title} {{Fundamentals of Nanomechanical Resonators}}}},\ \bibinfo {edition}
  {1st}\ ed.\ (\bibinfo  {publisher} {Springer International Publishing},\
  \bibinfo {year} {2016})\BibitemShut {NoStop}%
\bibitem [{\citenamefont {Verbridge}\ \emph {et~al.}(2006)\citenamefont
  {Verbridge}, \citenamefont {Parpia}, \citenamefont {Reichenbach},
  \citenamefont {Bellan},\ and\ \citenamefont {Craighead}}]{verbridge2006high}%
  \BibitemOpen
  \bibfield  {author} {\bibinfo {author} {\bibfnamefont {S.~S.}\ \bibnamefont
  {Verbridge}}, \bibinfo {author} {\bibfnamefont {J.~M.}\ \bibnamefont
  {Parpia}}, \bibinfo {author} {\bibfnamefont {R.~B.}\ \bibnamefont
  {Reichenbach}}, \bibinfo {author} {\bibfnamefont {L.~M.}\ \bibnamefont
  {Bellan}}, \ and\ \bibinfo {author} {\bibfnamefont {H.~G.}\ \bibnamefont
  {Craighead}},\ }\href@noop {} {\bibfield  {journal} {\bibinfo  {journal}
  {Journal of Applied Physics}\ }\textbf {\bibinfo {volume} {99}},\ \bibinfo
  {pages} {124304} (\bibinfo {year} {2006})}\BibitemShut {NoStop}%
\bibitem [{\citenamefont {Zwickl}\ \emph {et~al.}(2008)\citenamefont {Zwickl},
  \citenamefont {Shanks}, \citenamefont {Jayich}, \citenamefont {Yang},
  \citenamefont {Jayich}, \citenamefont {Thompson},\ and\ \citenamefont
  {Harris}}]{zwickl2008high}%
  \BibitemOpen
  \bibfield  {author} {\bibinfo {author} {\bibfnamefont {B.~M.}\ \bibnamefont
  {Zwickl}}, \bibinfo {author} {\bibfnamefont {W.~E.}\ \bibnamefont {Shanks}},
  \bibinfo {author} {\bibfnamefont {A.~M.}\ \bibnamefont {Jayich}}, \bibinfo
  {author} {\bibfnamefont {C.}~\bibnamefont {Yang}}, \bibinfo {author}
  {\bibfnamefont {B.}~\bibnamefont {Jayich}}, \bibinfo {author} {\bibfnamefont
  {J.~D.}\ \bibnamefont {Thompson}}, \ and\ \bibinfo {author} {\bibfnamefont
  {J.~G.~E.}\ \bibnamefont {Harris}},\ }\href@noop {} {\bibfield  {journal}
  {\bibinfo  {journal} {Applied Physics Letters}\ }\textbf {\bibinfo {volume}
  {92}},\ \bibinfo {pages} {103125} (\bibinfo {year} {2008})}\BibitemShut
  {NoStop}%
\bibitem [{\citenamefont {Verbridge}, \citenamefont {Craighead},\ and\
  \citenamefont {Parpia}(2008)}]{Verbridge2008}%
  \BibitemOpen
  \bibfield  {author} {\bibinfo {author} {\bibfnamefont {S.~S.}\ \bibnamefont
  {Verbridge}}, \bibinfo {author} {\bibfnamefont {H.~G.}\ \bibnamefont
  {Craighead}}, \ and\ \bibinfo {author} {\bibfnamefont {J.~M.}\ \bibnamefont
  {Parpia}},\ }\href {\doibase 10.1063/1.2822406} {\bibfield  {journal}
  {\bibinfo  {journal} {Applied Physics Letters}\ }\textbf {\bibinfo {volume}
  {92}},\ \bibinfo {pages} {013112} (\bibinfo {year} {2008})}\BibitemShut
  {NoStop}%
\bibitem [{\citenamefont {Chakram}\ \emph {et~al.}(2014)\citenamefont
  {Chakram}, \citenamefont {Patil}, \citenamefont {Chang},\ and\ \citenamefont
  {Vengalattore}}]{Chakram2014}%
  \BibitemOpen
  \bibfield  {author} {\bibinfo {author} {\bibfnamefont {S.}~\bibnamefont
  {Chakram}}, \bibinfo {author} {\bibfnamefont {Y.~S.}\ \bibnamefont {Patil}},
  \bibinfo {author} {\bibfnamefont {L.}~\bibnamefont {Chang}}, \ and\ \bibinfo
  {author} {\bibfnamefont {M.}~\bibnamefont {Vengalattore}},\ }\href {\doibase
  10.1103/PhysRevLett.112.127201} {\bibfield  {journal} {\bibinfo  {journal}
  {Physical Review Letters}\ }\textbf {\bibinfo {volume} {112}},\ \bibinfo
  {pages} {127201} (\bibinfo {year} {2014})}\BibitemShut {NoStop}%
\bibitem [{\citenamefont {Purdy}, \citenamefont {Peterson},\ and\ \citenamefont
  {Regal}(2013)}]{Purdy2013}%
  \BibitemOpen
  \bibfield  {author} {\bibinfo {author} {\bibfnamefont {T.~P.}\ \bibnamefont
  {Purdy}}, \bibinfo {author} {\bibfnamefont {R.~W.}\ \bibnamefont {Peterson}},
  \ and\ \bibinfo {author} {\bibfnamefont {C.~A.}\ \bibnamefont {Regal}},\
  }\href {\doibase 10.1126/science.1231282} {\bibfield  {journal} {\bibinfo
  {journal} {Science}\ }\textbf {\bibinfo {volume} {339}},\ \bibinfo {pages}
  {801} (\bibinfo {year} {2013})}\BibitemShut {NoStop}%
\bibitem [{\citenamefont {Wilson}(2012)}]{Wilson2012}%
  \BibitemOpen
  \bibfield  {author} {\bibinfo {author} {\bibfnamefont {D.~J.}\ \bibnamefont
  {Wilson}},\ }\emph {\bibinfo {title} {{Cavity Optomechanics with High-Stress
  Silicon Nitride Films}}},\ \href@noop {} {Ph.D. thesis},\ \bibinfo  {school}
  {California Institute of Technology} (\bibinfo {year} {2012})\BibitemShut
  {NoStop}%
\bibitem [{\citenamefont {Gavartin}, \citenamefont {Verlot},\ and\
  \citenamefont {Kippenberg}(2012)}]{Gavartin2012}%
  \BibitemOpen
  \bibfield  {author} {\bibinfo {author} {\bibfnamefont {E.}~\bibnamefont
  {Gavartin}}, \bibinfo {author} {\bibfnamefont {P.}~\bibnamefont {Verlot}},
\
  and\ \bibinfo {author} {\bibfnamefont {T.~J.}\ \bibnamefont {Kippenberg}},\
  }\href {\doibase 10.1038/nnano.2012.97} {\bibfield  {journal} {\bibinfo
  {journal} {Nature Nanotechnology}\ }\textbf {\bibinfo {volume} {7}},\
  \bibinfo {pages} {509} (\bibinfo {year} {2012})}\BibitemShut {NoStop}%
\bibitem [{\citenamefont {Bagci}\ \emph {et~al.}(2014)\citenamefont {Bagci},
  \citenamefont {Simonsen}, \citenamefont {Schmid}, \citenamefont {Villanueva},
  \citenamefont {Zeuthen}, \citenamefont {Appel}, \citenamefont {Taylor},
  \citenamefont {S{\o}rensen}, \citenamefont {Usami}, \citenamefont
  {Schliesser},\ and\ \citenamefont {Polzik}}]{Bagci2014}%
  \BibitemOpen
  \bibfield  {author} {\bibinfo {author} {\bibfnamefont {T.}~\bibnamefont
  {Bagci}}, \bibinfo {author} {\bibfnamefont {A.}~\bibnamefont {Simonsen}},
  \bibinfo {author} {\bibfnamefont {S.}~\bibnamefont {Schmid}}, \bibinfo
  {author} {\bibfnamefont {L.~G.}\ \bibnamefont {Villanueva}}, \bibinfo
  {author} {\bibfnamefont {E.}~\bibnamefont {Zeuthen}}, \bibinfo {author}
  {\bibfnamefont {J.}~\bibnamefont {Appel}}, \bibinfo {author} {\bibfnamefont
  {J.~M.}\ \bibnamefont {Taylor}}, \bibinfo {author} {\bibfnamefont
  {A.}~\bibnamefont {S{\o}rensen}}, \bibinfo {author} {\bibfnamefont
  {K.}~\bibnamefont {Usami}}, \bibinfo {author} {\bibfnamefont
  {A.}~\bibnamefont {Schliesser}}, \ and\ \bibinfo {author} {\bibfnamefont
  {E.~S.}\ \bibnamefont {Polzik}},\ }\href {\doibase 10.1038/nature13029}
  {\bibfield  {journal} {\bibinfo  {journal} {Nature}\ }\textbf {\bibinfo
  {volume} {507}},\ \bibinfo {pages} {81} (\bibinfo {year} {2014})}\BibitemShut
  {NoStop}%
\bibitem [{\citenamefont {Thompson}\ \emph {et~al.}(2008)\citenamefont
  {Thompson}, \citenamefont {Zwickl}, \citenamefont {Jayich}, \citenamefont
  {Marquardt}, \citenamefont {Girvin},\ and\ \citenamefont
  {Harris}}]{thompson2008strong}%
  \BibitemOpen
  \bibfield  {author} {\bibinfo {author} {\bibfnamefont {J.~D.}\ \bibnamefont
  {Thompson}}, \bibinfo {author} {\bibfnamefont {B.~M.}\ \bibnamefont
  {Zwickl}}, \bibinfo {author} {\bibfnamefont {A.~M.}\ \bibnamefont {Jayich}},
  \bibinfo {author} {\bibfnamefont {F.}~\bibnamefont {Marquardt}}, \bibinfo
  {author} {\bibfnamefont {S.~M.}\ \bibnamefont {Girvin}}, \ and\ \bibinfo
  {author} {\bibfnamefont {J.~G.~E.}\ \bibnamefont {Harris}},\ }\href@noop
{}
  {\bibfield  {journal} {\bibinfo  {journal} {Nature}\ }\textbf {\bibinfo
  {volume} {452}},\ \bibinfo {pages} {72} (\bibinfo {year} {2008})}\BibitemShut
  {NoStop}%
\bibitem [{\citenamefont {Brawley}\ \emph {et~al.}(2016)\citenamefont
  {Brawley}, \citenamefont {Vanner}, \citenamefont {Larsen}, \citenamefont
  {Schmid}, \citenamefont {Boisen},\ and\ \citenamefont
  {Bowen}}]{Brawley2014a}%
  \BibitemOpen
  \bibfield  {author} {\bibinfo {author} {\bibfnamefont {G.~a.}\ \bibnamefont
  {Brawley}}, \bibinfo {author} {\bibfnamefont {M.~R.}\ \bibnamefont {Vanner}},
  \bibinfo {author} {\bibfnamefont {P.~E.}\ \bibnamefont {Larsen}}, \bibinfo
  {author} {\bibfnamefont {S.}~\bibnamefont {Schmid}}, \bibinfo {author}
  {\bibfnamefont {A.}~\bibnamefont {Boisen}}, \ and\ \bibinfo {author}
  {\bibfnamefont {W.~P.}\ \bibnamefont {Bowen}},\ }\href {\doibase
  10.1038/ncomms10988} {\bibfield  {journal} {\bibinfo  {journal} {Nature
  Communications}\ }\textbf {\bibinfo {volume} {7}} (\bibinfo {year} {2016}),\
  10.1038/ncomms10988},\ \Eprint {http://arxiv.org/abs/1404.5746}
  {arXiv:1404.5746} \BibitemShut {NoStop}%
\bibitem [{\citenamefont {Tsaturyan}\ \emph {et~al.}(2016)\citenamefont
  {Tsaturyan}, \citenamefont {Barg}, \citenamefont {Polzik},\ and\
  \citenamefont {Schliesser}}]{Tsaturyan2016}%
  \BibitemOpen
  \bibfield  {author} {\bibinfo {author} {\bibfnamefont {Y.}~\bibnamefont
  {Tsaturyan}}, \bibinfo {author} {\bibfnamefont {A.}~\bibnamefont {Barg}},
  \bibinfo {author} {\bibfnamefont {E.~S.}\ \bibnamefont {Polzik}}, \ and\
  \bibinfo {author} {\bibfnamefont {A.}~\bibnamefont {Schliesser}},\ }\href
  {http://arxiv.org/abs/1608.00937} {\ ,\ \bibinfo {pages} {1} (\bibinfo
{year}
  {2016})},\ \Eprint {http://arxiv.org/abs/1608.00937} {arXiv:1608.00937}
  \BibitemShut {NoStop}%
\bibitem [{\citenamefont {Ghadimi}, \citenamefont {Wilson},\ and\ \citenamefont
  {Kippenberg}(2017)}]{Ghadimi2016}%
  \BibitemOpen
  \bibfield  {author} {\bibinfo {author} {\bibfnamefont {A.~H.}\ \bibnamefont
  {Ghadimi}}, \bibinfo {author} {\bibfnamefont {D.~J.}\ \bibnamefont {Wilson}},
  \ and\ \bibinfo {author} {\bibfnamefont {T.~J.}\ \bibnamefont {Kippenberg}},\
  }\href {\doibase 10.1021/acs.nanolett.7b00573} {\bibfield  {journal}
  {\bibinfo  {journal} {Nano Letters}\ }\textbf {\bibinfo {volume} {ASAP}}
  (\bibinfo {year} {2017}),\ 10.1021/acs.nanolett.7b00573},\ \Eprint
  {http://arxiv.org/abs/1603.01605} {arXiv:1603.01605} \BibitemShut {NoStop}%
\bibitem [{\citenamefont {Norte}, \citenamefont {Moura},\ and\ \citenamefont
  {Gr{\"{o}}blacher}(2016)}]{Norte2015}%
  \BibitemOpen
  \bibfield  {author} {\bibinfo {author} {\bibfnamefont {R.~A.}\ \bibnamefont
  {Norte}}, \bibinfo {author} {\bibfnamefont {J.~P.}\ \bibnamefont {Moura}},
\
  and\ \bibinfo {author} {\bibfnamefont {S.}~\bibnamefont {Gr{\"{o}}blacher}},\
  }\href {\doibase 10.1103/PhysRevLett.116.147202} {\bibfield  {journal}
  {\bibinfo  {journal} {Physical Review Letters}\ }\textbf {\bibinfo {volume}
  {116}},\ \bibinfo {pages} {1} (\bibinfo {year} {2016})},\ \Eprint
  {http://arxiv.org/abs/1511.06235} {arXiv:1511.06235} \BibitemShut {NoStop}%
\bibitem [{\citenamefont {Kurek}\ \emph {et~al.}(2017)\citenamefont {Kurek},
  \citenamefont {Carnoy}, \citenamefont {Larsen}, \citenamefont {Nielsen},
  \citenamefont {Hansen}, \citenamefont {Rades}, \citenamefont {Schmid},\
and\
  \citenamefont {Boisen}}]{Kurek2017}%
  \BibitemOpen
  \bibfield  {author} {\bibinfo {author} {\bibfnamefont {M.}~\bibnamefont
  {Kurek}}, \bibinfo {author} {\bibfnamefont {M.}~\bibnamefont {Carnoy}},
  \bibinfo {author} {\bibfnamefont {P.~E.}\ \bibnamefont {Larsen}}, \bibinfo
  {author} {\bibfnamefont {L.~H.}\ \bibnamefont {Nielsen}}, \bibinfo {author}
  {\bibfnamefont {O.}~\bibnamefont {Hansen}}, \bibinfo {author} {\bibfnamefont
  {T.}~\bibnamefont {Rades}}, \bibinfo {author} {\bibfnamefont
  {S.}~\bibnamefont {Schmid}}, \ and\ \bibinfo {author} {\bibfnamefont
  {A.}~\bibnamefont {Boisen}},\ }\href {\doibase 10.1002/anie.201700052}
  {\bibfield  {journal} {\bibinfo  {journal} {Angewandte Chemie International
  Edition}\ ,\ \bibinfo {pages} {3901}} (\bibinfo {year} {2017})}\BibitemShut
  {NoStop}%
\bibitem [{\citenamefont {Andersen}\ \emph {et~al.}(2016)\citenamefont
  {Andersen}, \citenamefont {Yamada}, \citenamefont {{Kumar E.K.}},
  \citenamefont {Andresen}, \citenamefont {Boisen},\ and\ \citenamefont
  {Schmid}}]{Andersen2016}%
  \BibitemOpen
  \bibfield  {author} {\bibinfo {author} {\bibfnamefont {A.~J.}\ \bibnamefont
  {Andersen}}, \bibinfo {author} {\bibfnamefont {S.}~\bibnamefont {Yamada}},
  \bibinfo {author} {\bibfnamefont {P.}~\bibnamefont {{Kumar E.K.}}}, \bibinfo
  {author} {\bibfnamefont {T.~L.}\ \bibnamefont {Andresen}}, \bibinfo {author}
  {\bibfnamefont {A.}~\bibnamefont {Boisen}}, \ and\ \bibinfo {author}
  {\bibfnamefont {S.}~\bibnamefont {Schmid}},\ }\href {\doibase
  10.1016/j.snb.2016.04.002} {\bibfield  {journal} {\bibinfo  {journal}
  {Sensors and Actuators B: Chemical}\ }\textbf {\bibinfo {volume} {233}},\
  \bibinfo {pages} {667} (\bibinfo {year} {2016})}\BibitemShut {NoStop}%
\bibitem [{\citenamefont {Yamada}\ \emph {et~al.}(2013)\citenamefont {Yamada},
  \citenamefont {Schmid}, \citenamefont {Larsen}, \citenamefont {Hansen},\
and\
  \citenamefont {Boisen}}]{Yamada2013b}%
  \BibitemOpen
  \bibfield  {author} {\bibinfo {author} {\bibfnamefont {S.}~\bibnamefont
  {Yamada}}, \bibinfo {author} {\bibfnamefont {S.}~\bibnamefont {Schmid}},
  \bibinfo {author} {\bibfnamefont {T.}~\bibnamefont {Larsen}}, \bibinfo
  {author} {\bibfnamefont {O.}~\bibnamefont {Hansen}}, \ and\ \bibinfo {author}
  {\bibfnamefont {A.}~\bibnamefont {Boisen}},\ }\href {\doibase
  10.1021/ac402585e} {\bibfield  {journal} {\bibinfo  {journal} {Analytical
  Chemistry}\ }\textbf {\bibinfo {volume} {85}},\ \bibinfo {pages} {10531}
  (\bibinfo {year} {2013})}\BibitemShut {NoStop}%
\bibitem [{\citenamefont {Biswas}\ \emph {et~al.}(2014)\citenamefont {Biswas},
  \citenamefont {Miriyala}, \citenamefont {Doolin}, \citenamefont {Liu},
  \citenamefont {Thundat},\ and\ \citenamefont {Davis}}]{Biswas2014}%
  \BibitemOpen
  \bibfield  {author} {\bibinfo {author} {\bibfnamefont {T.~S.}\ \bibnamefont
  {Biswas}}, \bibinfo {author} {\bibfnamefont {N.}~\bibnamefont {Miriyala}},
  \bibinfo {author} {\bibfnamefont {C.}~\bibnamefont {Doolin}}, \bibinfo
  {author} {\bibfnamefont {X.}~\bibnamefont {Liu}}, \bibinfo {author}
  {\bibfnamefont {T.}~\bibnamefont {Thundat}}, \ and\ \bibinfo {author}
  {\bibfnamefont {J.~P.}\ \bibnamefont {Davis}},\ }\href@noop {} {\bibfield
  {journal} {\bibinfo  {journal} {Analytical chemistry}\ }\textbf {\bibinfo
  {volume} {86}},\ \bibinfo {pages} {11368} (\bibinfo {year}
  {2014})}\BibitemShut {NoStop}%
\bibitem [{\citenamefont {Larsen}\ \emph {et~al.}(2013)\citenamefont {Larsen},
  \citenamefont {Schmid}, \citenamefont {Villanueva},\ and\ \citenamefont
  {Boisen}}]{Larsen2013}%
  \BibitemOpen
  \bibfield  {author} {\bibinfo {author} {\bibfnamefont {T.}~\bibnamefont
  {Larsen}}, \bibinfo {author} {\bibfnamefont {S.}~\bibnamefont {Schmid}},
  \bibinfo {author} {\bibfnamefont {L.~G.}\ \bibnamefont {Villanueva}}, \
and\
  \bibinfo {author} {\bibfnamefont {A.}~\bibnamefont {Boisen}},\ }\href
  {\doibase 10.1021/nn402057f} {\bibfield  {journal} {\bibinfo  {journal}
{ACS
  Nano}\ }\textbf {\bibinfo {volume} {7}},\ \bibinfo {pages} {6188} (\bibinfo
  {year} {2013})}\BibitemShut {NoStop}%
\bibitem [{\citenamefont {Ligenza}(1965)}]{Ligenza1965}%
  \BibitemOpen
  \bibfield  {author} {\bibinfo {author} {\bibfnamefont {J.~R.}\ \bibnamefont
  {Ligenza}},\ }\href {\doibase 10.1063/1.1714565} {\bibfield  {journal}
  {\bibinfo  {journal} {Journal of Applied Physics}\ }\textbf {\bibinfo
  {volume} {36}},\ \bibinfo {pages} {2703} (\bibinfo {year}
  {1965})}\BibitemShut {NoStop}%
\bibitem [{\citenamefont {Kraitchman}(1967)}]{Kraitchman1967}%
  \BibitemOpen
  \bibfield  {author} {\bibinfo {author} {\bibfnamefont {J.}~\bibnamefont
  {Kraitchman}},\ }\href {\doibase 10.1063/1.1709122} {\bibfield  {journal}
  {\bibinfo  {journal} {Journal of Applied Physics}\ }\textbf {\bibinfo
  {volume} {38}},\ \bibinfo {pages} {4323} (\bibinfo {year}
  {1967})}\BibitemShut {NoStop}%
\bibitem [{\citenamefont {Pulfrey}\ and\ \citenamefont
  {Reche}(1974)}]{Pulfrey1974}%
  \BibitemOpen
  \bibfield  {author} {\bibinfo {author} {\bibfnamefont {D.}~\bibnamefont
  {Pulfrey}}\ and\ \bibinfo {author} {\bibfnamefont {J.}~\bibnamefont
  {Reche}},\ }\href {\doibase 10.1016/0038-1101(74)90184-1} {\enquote {\bibinfo
  {title} {{Preparation and properties of plasma-anodized silicon dioxide
  films}},}\ } (\bibinfo {year} {1974})\BibitemShut {NoStop}%
\bibitem [{\citenamefont {Taylor}\ \emph {et~al.}(1988)\citenamefont {Taylor},
  \citenamefont {Eccleston}, \citenamefont {Ringnalda}, \citenamefont {Maher},
  \citenamefont {Eaglesham}, \citenamefont {Humphreys},\ and\ \citenamefont
  {Godfrey}}]{Taylor1988}%
  \BibitemOpen
  \bibfield  {author} {\bibinfo {author} {\bibfnamefont {S.}~\bibnamefont
  {Taylor}}, \bibinfo {author} {\bibfnamefont {W.}~\bibnamefont {Eccleston}},
  \bibinfo {author} {\bibfnamefont {J.}~\bibnamefont {Ringnalda}}, \bibinfo
  {author} {\bibfnamefont {D.~M.}\ \bibnamefont {Maher}}, \bibinfo {author}
  {\bibfnamefont {D.~J.}\ \bibnamefont {Eaglesham}}, \bibinfo {author}
  {\bibfnamefont {C.~J.}\ \bibnamefont {Humphreys}}, \ and\ \bibinfo {author}
  {\bibfnamefont {D.~J.}\ \bibnamefont {Godfrey}},\ }\href@noop {} {\bibfield
  {journal} {\bibinfo  {journal} {Journal de Physique}\ }\textbf {\bibinfo
  {volume} {49}},\ \bibinfo {pages} {393} (\bibinfo {year} {1988})}\BibitemShut
  {NoStop}%
\bibitem [{\citenamefont {Kennedy}\ \emph {et~al.}(1995)\citenamefont
  {Kennedy}, \citenamefont {Taylor}, \citenamefont {Eccleston}, \citenamefont
  {Arnoldbik},\ and\ \citenamefont {Habraken}}]{Kennedy1995}%
  \BibitemOpen
  \bibfield  {author} {\bibinfo {author} {\bibfnamefont {G.~P.}\ \bibnamefont
  {Kennedy}}, \bibinfo {author} {\bibfnamefont {S.}~\bibnamefont {Taylor}},
  \bibinfo {author} {\bibfnamefont {W.}~\bibnamefont {Eccleston}}, \bibinfo
  {author} {\bibfnamefont {W.~M.}\ \bibnamefont {Arnoldbik}}, \ and\ \bibinfo
  {author} {\bibfnamefont {F.~H.~M.}\ \bibnamefont {Habraken}},\ }\href@noop
{}
  {\bibfield  {journal} {\bibinfo  {journal} {Microelectronic Engineering}\
  }\textbf {\bibinfo {volume} {28}},\ \bibinfo {pages} {141} (\bibinfo {year}
  {1995})}\BibitemShut {NoStop}%
\bibitem [{\citenamefont {Kennedy}, \citenamefont {Buiu},\ and\ \citenamefont
  {Taylor}(1999)}]{Kennedy1999}%
  \BibitemOpen
  \bibfield  {author} {\bibinfo {author} {\bibfnamefont {G.~P.}\ \bibnamefont
  {Kennedy}}, \bibinfo {author} {\bibfnamefont {O.}~\bibnamefont {Buiu}},
\
  and\ \bibinfo {author} {\bibfnamefont {S.}~\bibnamefont {Taylor}},\ }\href
  {\doibase 10.1063/1.369678} {\bibfield  {journal} {\bibinfo  {journal}
  {Journal of Applied Physics}\ }\textbf {\bibinfo {volume} {85}},\ \bibinfo
  {pages} {3319} (\bibinfo {year} {1999})}\BibitemShut {NoStop}%
\bibitem [{\citenamefont {Smekal}, \citenamefont {Werner},\ and\ \citenamefont
  {Powell}(2005)}]{Smekal2005}%
  \BibitemOpen
  \bibfield  {author} {\bibinfo {author} {\bibfnamefont {W.}~\bibnamefont
  {Smekal}}, \bibinfo {author} {\bibfnamefont {W.~S.~M.}\ \bibnamefont
  {Werner}}, \ and\ \bibinfo {author} {\bibfnamefont {C.~J.}\ \bibnamefont
  {Powell}},\ }\href {\doibase 10.1002/sia.2097} {\bibfield  {journal}
  {\bibinfo  {journal} {Surface and Interface Analysis}\ }\textbf {\bibinfo
  {volume} {37}},\ \bibinfo {pages} {1059} (\bibinfo {year}
  {2005})}\BibitemShut {NoStop}%
\bibitem [{\citenamefont {Buiu}\ \emph {et~al.}(1998)\citenamefont {Buiu},
  \citenamefont {Kennedy}, \citenamefont {Gartner},\ and\ \citenamefont
  {Taylor}}]{Buiu1998}%
  \BibitemOpen
  \bibfield  {author} {\bibinfo {author} {\bibfnamefont {O.}~\bibnamefont
  {Buiu}}, \bibinfo {author} {\bibfnamefont {G.}~\bibnamefont {Kennedy}},
  \bibinfo {author} {\bibfnamefont {M.}~\bibnamefont {Gartner}}, \ and\
  \bibinfo {author} {\bibfnamefont {S.}~\bibnamefont {Taylor}},\ }\href
  {\doibase 10.1109/27.747889} {\bibfield  {journal} {\bibinfo  {journal}
{IEEE
  Transactions on Plasma Science}\ }\textbf {\bibinfo {volume} {26}},\ \bibinfo
  {pages} {1700} (\bibinfo {year} {1998})}\BibitemShut {NoStop}%
\bibitem [{\citenamefont {Sutardja}\ and\ \citenamefont
  {Oldham}(1989)}]{Sutardja1989}%
  \BibitemOpen
  \bibfield  {author} {\bibinfo {author} {\bibfnamefont {P.}~\bibnamefont
  {Sutardja}}\ and\ \bibinfo {author} {\bibfnamefont {W.}~\bibnamefont
  {Oldham}},\ }\href {\doibase 10.1109/16.43661} {\bibfield  {journal}
  {\bibinfo  {journal} {IEEE Transactions on Electron Devices}\ }\textbf
  {\bibinfo {volume} {36}},\ \bibinfo {pages} {2415} (\bibinfo {year}
  {1989})}\BibitemShut {NoStop}%
\bibitem [{\citenamefont {Kim}\ \emph {et~al.}(1996)\citenamefont {Kim},
  \citenamefont {An}, \citenamefont {Shin}, \citenamefont {Suh}, \citenamefont
  {Youn}, \citenamefont {Lee}, \citenamefont {Lee},\ and\ \citenamefont
  {Lee}}]{Kim1996}%
  \BibitemOpen
  \bibfield  {author} {\bibinfo {author} {\bibfnamefont {K.}~\bibnamefont
  {Kim}}, \bibinfo {author} {\bibfnamefont {M.~H.}\ \bibnamefont {An}},
  \bibinfo {author} {\bibfnamefont {Y.~G.}\ \bibnamefont {Shin}}, \bibinfo
  {author} {\bibfnamefont {M.~S.}\ \bibnamefont {Suh}}, \bibinfo {author}
  {\bibfnamefont {C.~J.}\ \bibnamefont {Youn}}, \bibinfo {author}
  {\bibfnamefont {Y.~H.}\ \bibnamefont {Lee}}, \bibinfo {author} {\bibfnamefont
  {K.~B.}\ \bibnamefont {Lee}}, \ and\ \bibinfo {author} {\bibfnamefont
  {H.~J.}\ \bibnamefont {Lee}},\ }\href@noop {} {\bibfield  {journal} {\bibinfo
   {journal} {Journal of Vacuum Science {\&} Technology B}\ }\textbf {\bibinfo
  {volume} {14}},\ \bibinfo {pages} {2667} (\bibinfo {year}
  {1996})}\BibitemShut {NoStop}%
\bibitem [{\citenamefont {Villanueva}\ and\ \citenamefont
  {Schmid}(2014)}]{Villanueva2014b}%
  \BibitemOpen
  \bibfield  {author} {\bibinfo {author} {\bibfnamefont {L.~G.}\ \bibnamefont
  {Villanueva}}\ and\ \bibinfo {author} {\bibfnamefont {S.}~\bibnamefont
  {Schmid}},\ }\href {\doibase 10.1103/PhysRevLett.113.227201} {\bibfield
  {journal} {\bibinfo  {journal} {Physical Review Letters}\ }\textbf {\bibinfo
  {volume} {113}},\ \bibinfo {pages} {1} (\bibinfo {year} {2014})},\ \Eprint
  {http://arxiv.org/abs/1405.6115} {arXiv:1405.6115} \BibitemShut {NoStop}%
\bibitem [{\citenamefont {Guo}\ and\ \citenamefont {Lal}(2003)}]{Guo2003}%
  \BibitemOpen
  \bibfield  {author} {\bibinfo {author} {\bibfnamefont {H.}~\bibnamefont
  {Guo}}\ and\ \bibinfo {author} {\bibfnamefont {A.}~\bibnamefont {Lal}},\
  }\href@noop {} {\bibfield  {journal} {\bibinfo  {journal} {Journal of
  Microelectromechanical Systems}\ }\textbf {\bibinfo {volume} {12}},\ \bibinfo
  {pages} {53} (\bibinfo {year} {2003})}\BibitemShut {NoStop}%
\bibitem [{\citenamefont {Southworth}\ \emph {et~al.}(2009)\citenamefont
  {Southworth}, \citenamefont {Barton}, \citenamefont {Verbridge},
  \citenamefont {Ilic}, \citenamefont {Fefferman}, \citenamefont {Craighead},\
  and\ \citenamefont {Parpia}}]{southworth2009stress}%
  \BibitemOpen
  \bibfield  {author} {\bibinfo {author} {\bibfnamefont {D.~R.}\ \bibnamefont
  {Southworth}}, \bibinfo {author} {\bibfnamefont {R.~A.}\ \bibnamefont
  {Barton}}, \bibinfo {author} {\bibfnamefont {S.~S.}\ \bibnamefont
  {Verbridge}}, \bibinfo {author} {\bibfnamefont {B.}~\bibnamefont {Ilic}},
  \bibinfo {author} {\bibfnamefont {A.~D.}\ \bibnamefont {Fefferman}}, \bibinfo
  {author} {\bibfnamefont {H.~G.}\ \bibnamefont {Craighead}}, \ and\ \bibinfo
  {author} {\bibfnamefont {J.~M.}\ \bibnamefont {Parpia}},\ }\href@noop {}
  {\bibfield  {journal} {\bibinfo  {journal} {Physical review letters}\
  }\textbf {\bibinfo {volume} {102}},\ \bibinfo {pages} {225503} (\bibinfo
  {year} {2009})}\BibitemShut {NoStop}%
\end{thebibliography}
\end{document}